# Beam loading analysis and control in standing wave cavities

Zheqiao Geng[1,†]
[1]Paul Scherrer Institut, 5232 Villigen PSI, Switzerland

**ABSTRACT**: The interaction between a particle beam and the accelerating mode of a radiofrequency (RF) cavity cause beam loading, representing the beam-induced cavity fields. Beam loading leads to amplitude and phase errors in the cavity fields and reduces the beam quality, especially in accelerators with large beam currents, wideband RF cavities, or circular machines where particles stay for multiple turns. Insight into the principle of beam loading is helpful to understand the beam measurement results and propose efficient compensation methods in low-level RF systems. In this work, the beam loading effects are studied with the equivalent circuit model of standing wave cavities. Analytical results of beam-induced cavity voltages are derived for both a single bunch and a bunch train using the phasor Laplace transform method. The results are general for wideband cavities with a bandwidth that may cover multiple harmonics of the bunch repetition frequency. Based on the analysis, control methods in form of feedforward and feedback are proposed to compensate for the beam loading. Simulation studies are carried out to validate these control methods with a cavity simulator including both the RF drive and beam loading. The analysis and control methods are also applicable to the beam in a circular accelerator with coupled-bunch instabilities, which are discussed in the last part of this paper. This work also acts as a supplementary material to another work of the author, in which the beam loading effects are analyzed only for narrow-band cavities with only one beam harmonic appearing in the cavity bandwidth.

## I. INTRODUCTION

Many accelerators employ standing-wave (SW) RF cavities to accelerate beams of charged particles. The electromagnetic fields in the cavity are built up with the RF drive power coupled via an input coupler. The particles interact with the cavity fields, either gaining energy or losing energy depending on the relative phase between the electric field and the particle arrival time. The beam loading, i.e., the beam-induced cavity fields, distorts the overall cavity fields modifying the accelerating voltage and phase felt by the beam. Typically, a low-level RF (LLRF) system is needed to regulate the cavity fields for achieving the desired beam parameters. Compensating for the beam loading effects by manipulating the amplitude and phase of the RF drive power is a crucial task of LLRF, which requires a deep understanding of the mechanisms of the interaction between the beam and cavity.

The topic of beam loading has been discussed in many articles [1-4]. The author's recent work [5] proposed a mathematical method, Phasor Laplace Transform (PLT), to model the RF cavities using transfer functions with complex coefficients. This method treats the input and output of a cavity as complex signals (phasors). The primary benefit is that the order of the cavity's differential equation is reduced from 2 to 1, simplifying the cavity response analysis significantly. With the PLT method, the beam induced cavity voltages of a single bunch and a bunch train have been derived in that work. However, the results are only suitable for narrow-band cavities covering only one harmonic of the bunch repetition frequency. Nowadays, many proton or heavy-ion accelerators [4, 6-8] adopt wideband cavities with a bandwidth covering multiple harmonics of the bunch repetition frequency. Furthermore, in circular accelerators, the beam current may contain rich frequency components if the buckets are filled with different charges (e.g., with empty buckets in the filling pattern) or the coupled-bunch instabilities are excited by factors like higher-order modes of the cavity [8-10]. All these scenarios motivate the analysis of beam loading effects for more general cases, e.g., in wideband cavities with arbitrary bunch patterns. The analysis leads to a systematic consideration of the beam loading compensation methods, resulting in a general controller architecture for LLRF systems. In this sense, this work can also be used as supplementary material for the book of reference [5].

Section II describes a general cavity model in the form of PLT. Section III derives the analytical results of the induced cavity voltages by a single bunch and an ideal bunch train with constant bunch charge and bunch-to-bunch time distance. Section IV proposes two beam loading compensation methods and validates them with simulation. Finally, Section V introduces how the results can be applied to circular accelerators with uneven bucket filling patterns or instabilities.

## II. CAVITY MODEL

The fundamental accelerating mode of a SW cavity can be modeled as a parallel resonance circuit [5], as shown in Fig. 1. The model input is a current phasor $\mathbf{i}_C = \mathbf{i}_{rf} + \mathbf{i}_b$, which is a superposition of the RF-induced drive current $\mathbf{i}_{rf}$ and the beam drive current $\mathbf{i}_b$. The output of the cavity model is a voltage phasor $\mathbf{v}_C$, representing the accelerating voltage and phase of the cavity field. The transfer function of the cavity in terms of Laplace transform can be written as

$$G(s) = \frac{2\omega_{1/2} R_L s}{s^2 + 2\omega_{1/2} s + \omega_0^2}$$

where $s$ is the complex frequency given by $s = \sigma + j\omega$ with real numbers $\sigma$ and $\omega$ and imaginary unit $j$. The cavity parameters, i.e., unloaded-lossless resonance frequency $\omega_0$, half bandwidth $\omega_{1/2}$, loaded resistance $R_L$, have been defined in Table 3.1 of the reference book [5] as $\omega_0 = 1/\sqrt{LC}$, $\omega_{1/2} = \omega_0/(2Q_L)$, $R_L = (R/Q)Q_L$, where $Q_L$ is the loaded quality factor and $R/Q$ is the normalized shunt impedance of the cavity (both are design parameters). To derive the cavity's Phasor Transfer Function (PTF), i.e., the relationship between the complex envelops of the cavity input and output, we shift the frequency part of $s$ by the RF carrier frequency $\omega_c$ (let $s = \hat{s} + j\omega_c$) and obtain

$$\begin{aligned}\tilde{\mathbf{G}}_C(\hat{s}) &= G(\hat{s} + j\omega_c) \\ &= \frac{\omega_{1/2} R_L g_h}{\hat{s} + \omega_{1/2} + j\Delta\omega_h} + \frac{\omega_{1/2} R_L g_l}{\hat{s} + \omega_{1/2} - j\Delta\omega_l}\end{aligned} \quad (1)$$

where

$$\begin{aligned}\omega_0' &= \sqrt{\omega_0^2 - \omega_{1/2}^2} \\ g_h &= 1 - j\,\omega_{1/2}/\omega_0', \quad \Delta\omega_h = \omega_0' + \omega_c, \\ g_l &= 1 + j\,\omega_{1/2}/\omega_0', \quad \Delta\omega_l = \omega_0' - \omega_c.\end{aligned} \quad (2)$$

---

† zheqiao.geng@psi.ch

The carrier frequency $\omega_c$ is typically chosen as the RF operating frequency of the cavity. Equation (1) splits the cavity transfer function into two passbands at a higher frequency $\Delta\omega_h$ and a lower frequency $\Delta\omega_l$ respectively. Shifting the frequency of $s$ is equivalent to mixing the RF signal with a local oscillator at $\omega_c$ and feeding the two resulting sidebands into the corresponding passbands of (1). The low-frequency passband is the cavity's dynamics for the input and output envelopes, while the results of the high-frequency passband are filtered out in practical LLRF systems. Therefore, from the viewpoint of RF control, only the low-frequency passband is of interest, and we write the cavity's PTF as

$$\mathbf{G}_C(\hat{s}) = \frac{2\omega_{1/2} R_L g_l}{\hat{s} + \omega_{1/2} - j\Delta\omega_l}. \quad (3)$$

The factor 2 is used to compensate for the gain loss when discarding the high-frequency passband.

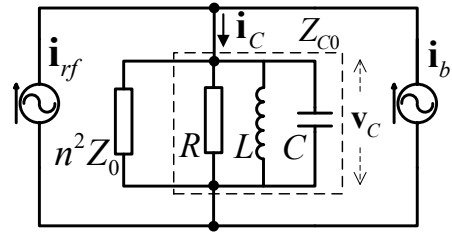

FIG 1. Circuit model of the accelerating mode of a SW cavity. Refer to Section 3.3.1 of the book [5].

To calculate the cavity response to a certain input, we derive the Laplace transform $X(s)$ of the input RF signal $x(t)$, and shift its frequency by assigning $s = \hat{s} + j\omega_c$ to obtain the PLT $\mathbf{X}(\hat{s}) = X(\hat{s} + j\omega_c)$, then the PLT of the output can be calculated as

$$\mathbf{Y}(\hat{s}) = \mathbf{G}_C(\hat{s})\mathbf{X}(\hat{s}). \quad (4)$$

The time-domain phasor $\mathbf{y}(t)$ of the output can be derived with the inverse Laplace transform, and the physical time-domain output $y(t)$ can be calculated by multiplying $e^{j\omega_c t}$ and taking the real part, written as

$$\begin{aligned}\mathbf{y}(t) &= \mathcal{L}^{-1}\{\mathbf{Y}(\hat{s})\}, \\ y(t) &= \operatorname{Re}\{\mathbf{y}(t)e^{j\omega_c t}\}.\end{aligned} \quad (5)$$

Note that we did not make any assumptions about the cavity bandwidth and resonance frequency. Therefore, the formulas above can be applied to the accelerating mode of any SW cavities.

### III. BEAM AND CAVITY INTERACTION

PLT is a powerful tool to analyze the cavity's response to different beam drive current $\mathbf{i}_b$. In the following analysis, the RF drive power, i.e., $\mathbf{i}_{rf}$, will be assigned to zero, and the cavity voltage is only induced by the beam.

#### A. Single bunch response

A single bunch with charge $Q_b$ is injected into the cavity at time $t = 0$. If the bunch duration is negligible compared to the observation time, the instant beam current can be written as $i_b(t) = Q_b\delta(t)$, where $\delta(t)$ is the Dirac delta function. The PLT of $i_b(t)$ is $\mathbf{I}_b(\hat{s}) = Q_b$, then the PLT of the induced cavity voltage can be calculated as

$$\begin{aligned}\mathbf{V}_{Cbs}(\hat{s}) &= \mathbf{G}_C(\hat{s})\mathbf{I}_b(\hat{s}) \\ &= \frac{2\omega_{1/2} R_L Q_b g_l}{\hat{s} + \omega_{1/2} - j\Delta\omega_l}.\end{aligned} \quad (6)$$

The time-domain phasor of the cavity voltage is

$$\mathbf{v}_{Cbs}(t) = 2\omega_{1/2} R_L Q_b g_l e^{-(\omega_{1/2} - j\Delta\omega_l)t}. \quad (7)$$

And the physical instant voltage in the cavity can be calculated as

$$v_{Cbs}(t) = \text{Re}\{\mathbf{v}_{Cbs}(t)e^{j\omega_c t}\}$$
$$= 2\omega_{1/2}R_L Q_b e^{-\omega_{1/2}t}\left(\cos\omega'_0 t - \frac{\omega_{1/2}}{\omega'_0}\sin\omega'_0 t\right) \quad (8)$$

Note that for a narrow-band cavity, we have $\omega_{1/2} \ll \omega_c$, $|\hat{s}| \ll \omega_c$, $\omega_c \approx \omega_0$, and the second term in the bracket of (8) is negligible.

We will demonstrate the beam loading effects and the controls with simulation for a test cavity referring to the CEPC (circular electron positron collider) design [8]. The cavity parameters used in simulation are listed below:

- Unloaded and lossless resonance frequency $\omega_0 = 2\pi f_0 = 2\pi \times 650\text{e}6$ rad/s.
- Normalized shunt impedance $R/Q = 106.5$ Ω.
- Loaded quality factor $Q_L = 1.5\text{e}5$.
- RF control loop delay $\tau = 1$ μs.
- Bunch charge $Q_b = 2.243\text{e}{-08}$ C.
- Bunch repetition frequency $\omega_b = \omega_0 / 21682$.

These parameters may not reflect the actual operating conditions of the CEPC cavity. They are chosen to demonstrate the beam loading effects of a bunch train (see next subsection) with multiple harmonics falling into the cavity bandwidth.

The parameters in the formulas above can be derived with the cavity and beam parameters: $\omega_{1/2} = \omega_0/(2Q_L)$, $R_L = (R/Q)Q_L$, and $g_l$ and $\Delta\omega_l$ are given by Eq. (2). We define the carrier frequency $\omega_c = \omega_0$. The cavity responses of a single bunch injected at different times in terms of amplitudes and phases w.r.t the carrier frequency is depicted in Fig. 8.

### B. Ideal bunch train response

Figure 2 is the instant beam current of an ideal bunch train starting from $t = -\infty$ and lasting until $t = +\infty$. The bunch charge is $Q_b$ and the time interval between adjacent bunches is $T_b$. One of the bunches arrives at time 0. We will calculate the cavity voltage induced by such a beam.

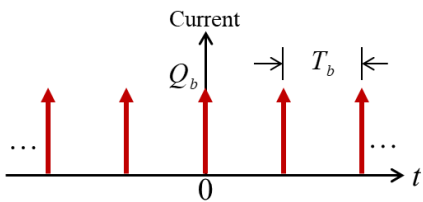

FIG 2. Instant beam current of an ideal bunch train.

The instant beam current can be presented as
$$i_b(t) = Q_b \sum_{n=-\infty}^{\infty} \delta(t - nT_b). \quad (9)$$

We consider it in the frequency domain. Its Fourier series can be written as
$$i_b(t) = \frac{A_0}{2} + \sum_{k=1}^{\infty}[A_k\cos(k\omega_b t) + B_k\sin(k\omega_b t)] \quad (10)$$

where the bunch repetition frequency is defined as $\omega_b = 2\pi/T_b$. The coefficients $A_k$ ($k = 0, 1, 2, \ldots$) and $B_k$ ($k = 1, 2, \ldots$) are calculated as

$$A_k = \frac{2Q_b}{T_b} \int_{-T_b/2}^{T_b/2} \sum_{n=-\infty}^{\infty}\delta(t-nT_b)\cos(k\omega_b t)\,dt$$
$$= \frac{2Q_b}{T_b} = 2I_{b0}, \quad (11)$$

$$B_k = \frac{2Q_b}{T_b} \int_{-T_b/2}^{T_b/2} \sum_{n=-\infty}^{\infty}\delta(t-nT_b)\sin(k\omega_b t)\,dt$$
$$= 0.$$

The average beam current is defined as $I_{b0} = Q_b/T_b$. Therefore, we can write the instant beam current starting from $t = 0$ as

$$i_b(t) = \left[I_{b0} + 2I_{b0}\sum_{k=1}^{\infty}\cos(k\omega_b t)\right]u(t) \quad (12)$$

where $u(t)$ is a unit step function at $t = 0$. The beam current consists of a DC term that equals the average current and infinite harmonics of the bunch repetition frequency. The Laplace transform of (12) is

$$I_b(s) = \frac{I_{b0}}{s} + 2I_{b0}\sum_{k=1}^{\infty}\frac{s}{s^2 + k^2\omega_b^2} \quad (13)$$

and the PLT can be written as
$$\mathbf{I}_b(\hat{s}) = I_b(s)|_{s=\hat{s}+j\omega_c}$$
$$= \sum_{k=-\infty}^{\infty}\frac{I_{b0}}{\hat{s}+j(\omega_c-k\omega_b)}. \quad (14)$$

Then, the PLT of the beam induced cavity voltage can be calculated as
$$\mathbf{V}_{Cbt}^+(\hat{s}) = \mathbf{G}_C(\hat{s})\mathbf{I}_b(\hat{s})$$
$$= \sum_{k=-\infty}^{\infty}\frac{2\omega_{1/2}R_L g_l}{\hat{s}+\omega_{1/2}-j\Delta\omega_l}\frac{I_{b0}}{\hat{s}+j(\omega_c-k\omega_b)}$$
$$= \sum_{k=-\infty}^{\infty}\frac{a_k}{\hat{s}+\omega_{1/2}-j\Delta\omega_l} + \frac{b_k}{\hat{s}+j(\omega_c-k\omega_b)}. \quad (15)$$

The superscript "+" indicates that this cavity voltage is produced by the beam current at $t \geq 0$. We work out the coefficients as

$$b_k = -a_k = \frac{2\omega_{1/2}R_L I_{b0} g_l}{\omega_{1/2} - j(\omega'_0 - k\omega_b)}. \quad (16)$$

The time-domain phasor of (15) can be calculated with the inverse Laplace transform as

$$\mathbf{v}^{+}_{Cbt}(t) = \sum_{k=-\infty}^{\infty} b_k \left( e^{-j(\omega_c - k\omega_b)t} - e^{-(\omega_{1/2} - j\Delta\omega_l)t} \right). \quad (17)$$

The term $b_k e^{-(\omega_{1/2} - j\Delta\omega_l)t}$ is the cavity's transient response to each beam harmonic and vanishes with time increases. The term $b_k e^{-j(\omega_c - k\omega_b)t}$ that oscillates at the frequency $\omega_c - k\omega_b$ with a constant amplitude $b_k$ is the cavity's steady-state response to the $k$th beam harmonic.

The bunch train current is presented as a Fourier series, a sum of infinite number of cosine functions. We only take the series' value at $t \geq 0$, as Eq. (12). In practice, only finite number of harmonics of the bunch repetition frequency contributing significantly are considered when calculating the beam-induced cavity voltage. By summing up these selected harmonics, it is not possible to reconstruct the delta-function-like beam current as in Fig. 2. Instead, we obtain a narrow pulse around each bunch injection time, spreading the beam current of a bunch into a short time range. Therefore, when calculating the cavity response with (17) with finite number of harmonics, only half of the beam current pulse around $t = 0$ is included. See Fig. 3. To get the correct cavity response around $t = 0$, the response of a half bunch should be added to (17), resulting in the total response of a cavity for a bunch train as

$$\begin{aligned}\mathbf{v}_{Cbt}(t) &= \mathbf{v}^{+}_{Cbt}(t) + \mathbf{v}_{Cbs}(t)/2 \\ &= \sum_{k=-N}^{N} \frac{2\omega_{1/2} R_L I_{b0} g_l}{\omega_{1/2} - j(\omega'_0 - k\omega_b)} \left( e^{-j(\omega_c - k\omega_b)t} - e^{-(\omega_{1/2} - j\Delta\omega_l)t} \right) + \\ &\quad \omega_{1/2} R_L Q_b g_l e^{-(\omega_{1/2} - j\Delta\omega_l)t}\end{aligned} \quad (18)$$

with $N < \infty$. Here $\mathbf{v}_{Cbs}$ is the single-bunch response given in (7).

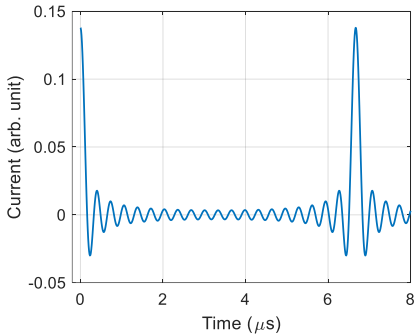

FIG 3. Reconstructed beam current waveform with 20 harmonics at $t \geq 0$. The beam repetition frequency here is chosen arbitrarily.

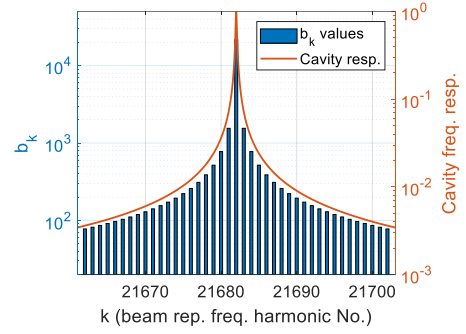

FIG 4. Values of $b_k$ for different $k$ values for the CEPC cavity. The cavity frequency response is also plotted.

Equation (16) implies that the oscillation amplitude $b_k$ in the beam-induced cavity voltage is independent of $\omega_c$ but determined by the frequency difference $\omega'_0 - k\omega_b$. Figure 4 shows the values of $b_k$ around the cavity resonance frequency for the CEPC cavity, with the same parameters given in Section III.A. It shows that only the beam harmonics with their frequencies fall into the cavity bandwidth will introduce significant beam loading in the cavity, and the magnitudes are determined by the cavity frequency response at those frequencies.

Equation (18) calculates the cavity response by summing up the responses of beam harmonics. Another method, which is more straightforward, calculates the beam-induced cavity voltage by adding the single-bunch response of each bunch at its injection time. We compare the results of these two methods using the CEPC cavity model. Its cavity voltage induced by the bunch train starting from $t = 0$ is simulated and the results are shown in Fig. 5. It shows that the two methods lead to the same result.

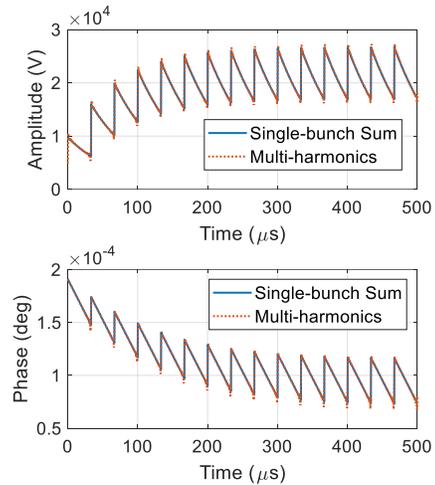

FIG 5. Bunch train induced cavity voltage. The curves of "Multi-harmonics" are calculated with (18) by summing up the responses of 41 beam harmonics around the carrier frequency $\omega_c$ (20 harmonics at each sideband and the harmonic at $\omega_c = 21682 \times \omega_b$). The curves of "Single-bunch Sum" are calculated by summing up the single-bunch response (7) of each bunch with a proper time delay.

Figure 6 shows the cavity responses to three selected beam harmonics. The cavity response to the 21682th harmonic is the fundamental beam loading as a baseband envelope of the carrier frequency. The first lower (21681th) and higher (21683th) harmonics produce outputs with phases changing linearly due to the frequency offset $k\omega_b - \omega_c$. Figure 7 is the spectrum of the overall cavity voltage induced by the bunch train, where the contributions from different beam harmonics are visible in the zoomed plot.

The results in this subsection have assumed a bunch train with constant bunch rate and bunch charge. For any bunch train with a repetitive pattern, the same method can be used to study its beam loading effects.

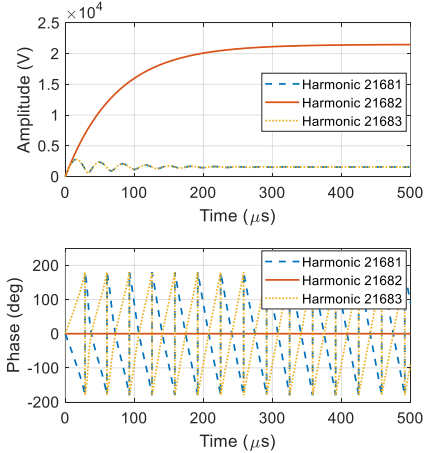

FIG 6. Responses of the cavity to selected beam harmonics.

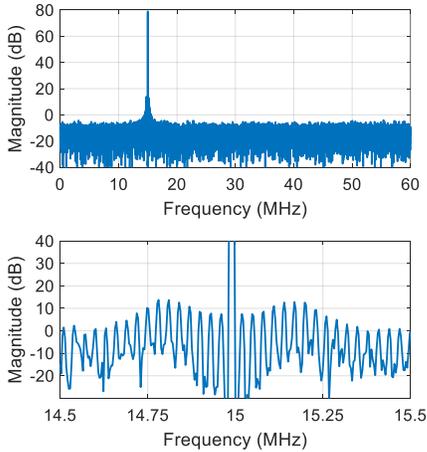

FIG 7. Spectrum of beam-induced cavity voltage down-converted to an intermediate frequency (IF) $f_{IF}$ = 14.99 MHz and sampled at $f_s$ = 119.92 MHz. The bunch repetition frequency is 29.98 kHz. Top: full spectrum; Bottom: zoom in around the IF frequency.

### C. Beam phase definition

To complete the theory of beam loading, we need to define the beam accelerating phase to represent the relative timing between the bunch arrival time and the cavity field oscillation with the RF drive power present.

Equations (7) and (18) are the cavity voltages induced by a single bunch and a bunch train with the first bunch injected at $t = 0$. If the bunches are delayed by $\Delta t$, the induced cavity voltage can be derived from the time shifting rule of the Laplace transform. When the system input $x(t)$ is delayed by $\Delta t$, its Laplace transform changes to be $X(s)e^{-s\Delta t}$, and the PLT becomes $\mathbf{X}(\hat{s})e^{-j\omega_c\Delta t}e^{-\hat{s}\Delta t}$. Therefore, according to (4), the output PLT becomes $\mathbf{Y}(\hat{s})e^{-j\omega_c\Delta t}e^{-\hat{s}\Delta t}$, which introduces to the output a carrier-frequency phase shift $-\omega_c\Delta t$ and an envelope delay $\Delta t$. We define $-\omega_c\Delta t$ as the phase of the beam drive current $\mathbf{i}_b$, and denote it as

$$\varphi_{b0} = -\omega_c \Delta t . \quad (19)$$

Then, after delaying the bunches by $\Delta t$, the cavity responses to a single bunch and a bunch train can be derived from (7) and (18) as

$$\begin{aligned}\mathbf{v}_{Cbs,delay}(t) &= \mathbf{v}_{Cbs}(t + \varphi_{b0}/\omega_c)e^{j\varphi_{b0}}, \\ \mathbf{v}_{Cbt,delay}(t) &= \mathbf{v}_{Cbt}(t + \varphi_{b0}/\omega_c)e^{j\varphi_{b0}}.\end{aligned} \quad (20)$$

These expressions are valid for $t + \varphi_{b0}/\omega_c \geq 0$. As an example, we delay the single bunch injection time to the CEPC cavity by 0.577 ns, corresponding to $\varphi_{b0} = -3\pi/4$, and compare the cavity responses. See Fig. 8 and Fig. 9.

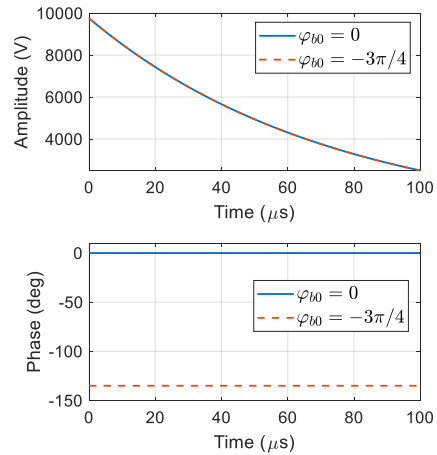

FIG 8. Cavity response amplitudes and phases for a single bunch injected at $t = 0$ and $t = 0.577$ ns.

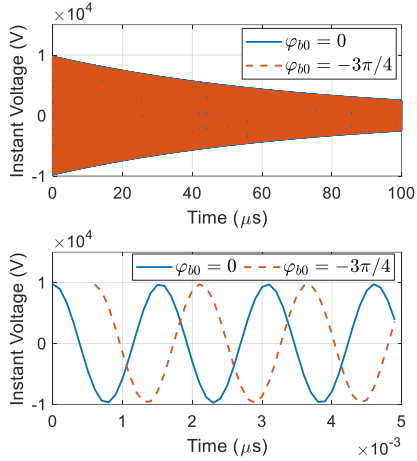

FIG 9. Cavity response physical signals for a single bunch injected at $t = 0$ and $t = 0.577$ ns. The bottom plot is the zoom in of the top one.

The overall cavity voltage is a superposition of the RF-induced and beam-induced cavity voltages, written in the phasor format as $\mathbf{v}_C = v_C e^{j\varphi_C} = \mathbf{v}_{Crf} + \mathbf{v}_{Cb}$. Here $v_C$ and $\varphi_C$ are the amplitude and phase of the overall cavity voltage, $\mathbf{v}_{Crf}$ is the RF-induced and $\mathbf{v}_{Cb}$ is the beam-induced cavity voltages. The beam accelerating phase (i.e., beam phase), defined to be zero for on-crest acceleration, is related to the phases of $\mathbf{v}_C$ and $\mathbf{i}_b$. Figure 5 is the cavity voltage induced by a beam with $\varphi_{b0} = 0$ (we neglect the small phase deviation in the plot). If the initial cavity voltage before injecting the beam has a phase $\varphi_C = 0$, then the beam induced voltage in Fig. 5 will increase the overall cavity voltage because they are in phase. It implies that the beam will lose energy in the cavity, and the beam phase is $\pi$, corresponding to a maximum deceleration of the beam. Therefore, the beam phase (denoted as $\varphi_b$) should be calculated as

$$\varphi_b = \pi + \varphi_C - \varphi_{b0}. \quad (21)$$

When the bunch arrival time is delayed by $\Delta t > 0$, the beam drive current phase $\varphi_{b0}$ obtains a negative change according to (19), then the beam phase changes in positive.

## IV. BEAM LOADING COMPENSATION

Cavities are typically controlled by LLRF systems, which manipulate the amplitude and phase of the RF drive power to regulate the cavity field for tracking the setpoint and suppressing the disturbances, including the beam loading. For wideband cavities, the standard Proportional Integral (PI) feedback controller may not be sufficient to compensate for the beam loading. The achievable control bandwidth of the PI controller is limited by the loop delay – the overall group delay across the feedback loop. Even if we shorten the distance between the LLRF controller and the cavity, it is difficult to reduce the loop delay within 1 μs. To compensate for the beam loading caused by multiple harmonics of the beam current in a wideband cavity, a closed-loop bandwidth much larger than the capability of the PI controller may be needed.

In addition to a PI controller, feedforward control can be introduced if the beam loading is stationary with constant or slowly changing patterns (slower than the adaptation speed of the feedforward controller). Random fluctuations in the beam loading, caused by nonrepetitive beam current or arrival time changes, should be compensated for by feedback. Special feedback can be applied to the beam harmonic frequencies to enhance the capability of the traditional PI feedback controller. We will discuss both algorithms.

A general RF controller is depicted in Fig. 10. The quantities are explained as follows: $\mathbf{v}_C$ is the phasor of the cavity voltage measurement, $\mathbf{v}_{C,SP}$ is the setpoint of the cavity voltage phasor, $\mathbf{v}_{err}$ is the error, $\mathbf{v}_{FB}$ is the feedback controller output, $\mathbf{v}_{FF}$ is the feedforward controller output and $\mathbf{v}_{act}$ is the total drive signal produced by the RF actuator to the amplifier.

To test the control algorithms, a cavity simulator is created. It is a continuously running software process simulating the behaviors of the cavity and the proposed feedforward/feedback controllers. The simulator implements the model of the CEPC cavity with the parameters given in Section III.A. The beam loading is introduced by applying the single-bunch result (7) at each bunch's injection time. The cavity voltage setpoint is 1 MV with a beam phase of -50 degrees. In simulation, the carrier frequency is chosen to be 650 MHz, the operating frequency of the CEPC cavity. To better view the cavity voltage spectrum, the carrier frequency is down converted to $f_{IF} = 14.99$ MHz. The sampling frequency of the RF controller is chosen to be $f_s = 119.92$ MHz. Figure 11 shows the graphical user interface of this cavity simulator.

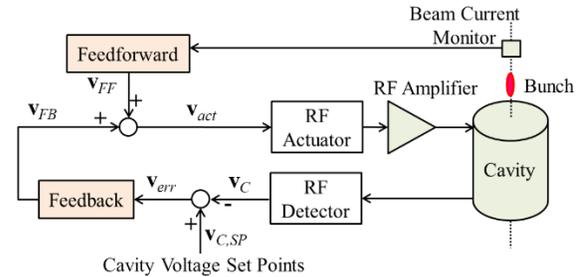

FIG 10. A general RF controller for SW cavities.

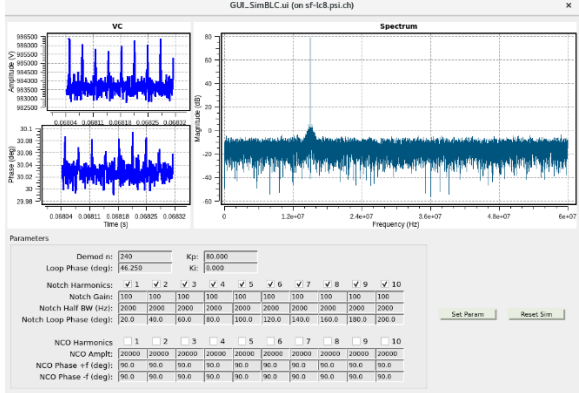

FIG 11. Panel for the cavity simulator. The feedforward and feedback controllers proposed in this section are implemented and configurable via the panel.

### A. Feedforward control

The beam current (12) consists of a DC component and harmonics of the bunch repetition frequency. Accordingly, we can consider the following feedforward signal in the RF drive to compensate for the beam loading:

$$\mathbf{v}_{FF} = A_{dc} e^{j\varphi_{dc}} + \sum_{k=1}^{N} A_k \left( e^{jk\omega_b} e^{j\varphi_k^+} + e^{-jk\omega_b} e^{j\varphi_k^-} \right). \quad (22)$$

The first term is to compensate for the DC beam loading and the harmonic terms for the beam loadings of $N$ beam harmonics. The beam harmonics at the upper sideband (with frequencies $k\omega_b$) and the lower sideband (with frequencies $-k\omega_b$) of the cavity are compensated for separately because they may require different rotation angles. The scale factors $A_{dc}$, $A_k$, and the rotation angles $\varphi_{dc}$, $\varphi_k^+$ and $\varphi_k^-$ should be determined empirically. The feedforward controller is depicted graphically in Fig. 12, where the I/Q control scheme has been adopted with $\mathbf{v}_{FF} = I_{FF} + jQ_{FF}$.

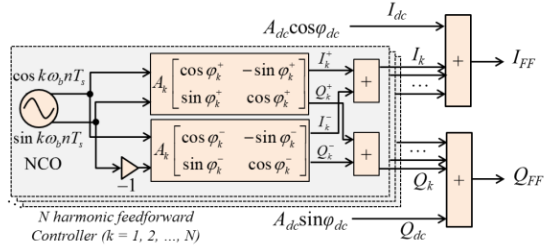

FIG 12. Feedforward controller for DC and harmonic beam loading compensation.

Numerically controlled oscillators (NCOs) are used to produce the beam harmonic frequencies. Compensating for the DC beam loading has been discussed in many articles [2, 11, 12]. Here we only focus on the beam loading compensation for the beam harmonics with frequencies falling into the cavity bandwidth.

For each harmonic beam loading, the feedforward compensation coefficients can be determined with the following procedure:

1. Set an initial value of $A_k$, which should be large enough to produce detectable changes in the cavity voltage spectrum at the corresponding beam harmonic frequencies.
2. Make a rough scan of $\varphi_k^+$ and $\varphi_k^-$ simultaneously for a full cycle with several steps. At each step, read the waveform of the cavity pickup signal, calculate its power spectral density (PSD), and record the PSD magnitudes at the frequencies of the $k$th beam harmonic at both sidebands.
3. Use the results in step 2 to determine the rough optimal values of $\varphi_k^+$ and $\varphi_k^-$ resulting in minimum PSDs at the corresponding frequencies.
4. Make a fine scan of $\varphi_k^+$ and $\varphi_k^-$ simultaneously around their rough optimal values respectively and record the PSDs at the corresponding frequencies.
5. Fit parabolic functions of the scan results in 4 and find the optimal values of $\varphi_k^+$ and $\varphi_k^-$ respectively. Set the optimal phases to the feedforward controller.
6. Use the scan method above (rough scan and then fine scan) to find the optimal value of $A_k$ and set it to the feedforward controller.

This method is applied to the cavity simulator for validation. Figure 13 shows the scan results to determine the compensation coefficients for one of the harmonics. Figure 14 is the cavity voltage spectrum with the beam loadings of 10 beam harmonics on each sideband compensated by the feedforward controller.

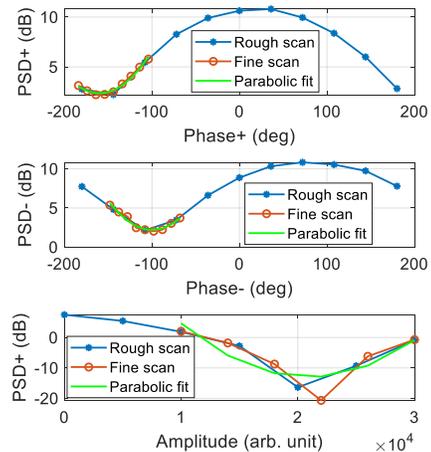

FIG 13. Optimization of the feedforward coefficients of a harmonic. "Phase+" stands for $\varphi_k^+$ and "Phase–" for $\varphi_k^-$; "Amplitude" stands for $A_k$; "PSD+" and "PSD–" are the PSDs at the beam harmonic frequencies at both sidebands.

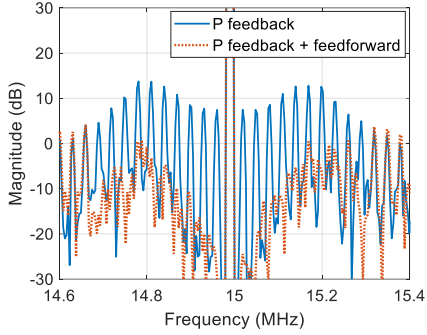

FIG 14. Spectrum of cavity voltage with the beam loadings compensated by the feedforward controller. The P feedback is used to keep a constant average cavity voltage. The spectrum is zoomed around the noise floor, and the carrier magnitude can be read in Fig. 7.

Feedforward is practical to compensate for repetitive beam loadings. When the system conditions change (e.g., with a different bunch repetition frequency, bunch charge, cavity voltage, beam phase, operating point of a nonlinear device, etc.), the compensation coefficients may need to be re-optimized. Specifically, if only the beam parameters drift slowly, non-invasive optimization algorithms [13], such as the random walk optimization, extremum seeking, or Bayesian optimization, can be used to track the drift by adapting the compensation coefficients.

In the feedforward implementation of Fig. 12, NCOs are used as the frequency sources of beam harmonics. In practice, one can also pick up the beam signal and derive the beam harmonic frequencies with the processing in Fig. 15 [6].

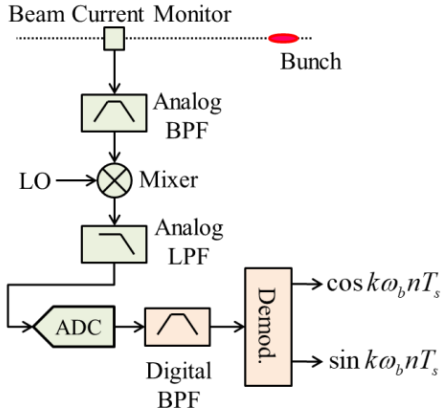

FIG 15. Derive the beam harmonic frequencies from the beam pickup signal to replace the NCOs. Note that the amplitude of the output is normalized to 1.

The beam current monitor is typically wideband to pick up the instant beam current signal. An analog bandpass filter (BPF) limits the band of the beam pickup signal around the cavity resonance frequency. The output contains the beam harmonics interacting with the cavity. We use a local oscillator (LO) (typically with a frequency of tens of MHz smaller than the RF operating frequency) to mix with the band limited beam signal, resulting in an IF signal and a signal with summed frequency. The IF frequency should be large enough to cover all the beam harmonics that interact with the cavity, but not too large to avoid being too sensitive to the ADC clock jitter. The analog low-pass filter (LPF) selects the IF frequency after the mixer. The IF signal is sampled by an ADC with a sampling rate typically several times larger than the IF frequency. At the digital side, the samples of the IF signal are bandpass filtered to pick up the frequency of $\omega_{IF} + k\omega_b$. The digital demodulation shifts the IF frequency to DC and derive a complex signal with the in-phase (I) part $\cos k\omega_b nT_s$ and the quadrature (Q) part $\sin k\omega_b nT_s$, just as the NCO output. The rest of the feedforward controller and the calibration algorithm remain the same. This implementation can track the beam drift with feedforward automatically. Another benefit is that when the beam is switched off, the feedforward signal is automatically switched off due to the loss of the beam pickup signal. With the implementation in Fig. 12, other means should be used to get the beam on/off status and switch on/off the feedforward signal correspondingly.

### B. Feedback control

Feedback is required to suppress random disturbances to the cavity fields, such as the time varying beam loading, cavity detuning, and RF amplifier fluctuations, etc. Most disturbances are at low frequencies and can be controlled by the traditional PI feedback, which increases the low-frequency open-loop gain and suppresses the low-frequency disturbances when closing the loop. With PI feedback control, the closed-loop bandwidth is limited by the loop delay, and the controller's capability of disturbance rejection at high frequencies is typically much weaker.

As seen in Fig. 7, the fluctuations of the beam-introduced cavity voltage spread into a wide frequency range and may not be effectively suppressed by the PI feedback control. We can consider increasing the open loop gain locally at the beam harmonic frequencies to better suppress the fluctuations at these frequencies in the cavity voltage. The proposed feedback controller is described as

$$\mathbf{K}(\hat{s}) = K_P + \frac{K_I}{\hat{s}} + \sum_{k=1}^{M} \left( \frac{g_k e^{j\theta_k} h_k}{\hat{s} + h_k - j\omega_{n,k}} + \frac{g_k e^{-j\theta_k} h_k}{\hat{s} + h_k + j\omega_{n,k}} \right). \quad (23)$$

The first two terms are PI feedback with $K_P$ and $K_I$ the proportional and integral gains respectively. The sum terms increase the gains at $\pm\omega_{n,k}$ offset from the carrier frequency and are denoted as "harmonic feedback". Note that if the disturbance frequency is not constant and evolves slowly (e.g., during the filling process of a storage ring equipped with harmonic cavities), the controller could be adapted with a time-varying $\omega_{n,k}$. Here we limit our discussion to the simple case with constant disturbance frequencies, i.e., to compensate for the beam loadings caused by the beam harmonics, for which we choose $\omega_{n,k} = k\omega_b$. The other parameters are explained as follows: $g_k$ is the harmonic feedback gain; $\theta_k$ is used to tune the phase response around the beam harmonic frequency and keep the closed loop stable; $h_k$ is the half bandwidth of the harmonic feedback. The sum terms are called "multi-harmonic feedback control" [6, 7]. The structure of the controller is shown in Fig. 16.

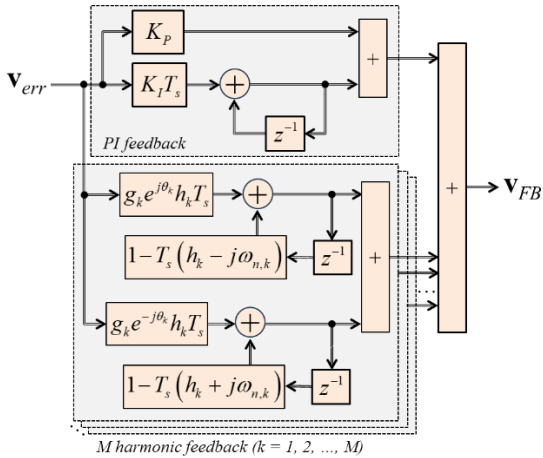

FIG 16. An RF controller with PI and multi-harmonic feedback controls. The signals between different blocks are phasors (i.e., I/Q control). Here $z^{-1}$ is a unit clock cycle delay in digital implementation.

We apply the controller (23) to the cavity model. The open-loop transfer function is given by $\mathbf{L}(\hat{s}) = \mathbf{G}_C(\hat{s})\mathbf{K}(\hat{s})$, whose frequency response determines the closed-loop stability and control capabilities. The gain margin (GM) of the open loop can be estimated to be $GM \approx \pi/(2\tau\omega_{1/2})$, where $\tau$ is the loop delay and $\omega_{1/2}$ is the cavity's half bandwidth. That is, GM is roughly the ratio between the frequency where the loop delay-caused phase lag reaches $\pi/2$ and the cavity half bandwidth. The estimation is based on two facts: 1. With only P feedback control, the closed-loop bandwidth is roughly $K_p$ times the cavity half bandwidth; 2. The cavity itself causes $-\pi/2$ phase lag at frequencies much larger than the cavity half bandwidth.

The CEPC cavity will be used to study the feedback control algorithms via simulation. With the parameters given in Section III.A, the gain margin is about 115, therefore, the PI feedback parameters are chosen as $K_p = 80$ and $K_I = 0$ (without integral control). For all harmonic feedback controllers, we choose a gain $g_k = 100$ and a half bandwidth $h_k = 2\pi \times 2000$ rad/s. In practice, we should implement as many harmonic feedback controllers as possible (maximize the $M$) to suppress the beam loading effects. However, the loop stability will limit the achievable maximum $M$. In the CEPC cavity example, if we set $\theta_k = 0°$, the maximum $M$ can only reach 5, and from the 6th harmonic feedback, the loop becomes unstable. As see in Fig. 17, the loop phases at the 6th beam harmonic frequencies at both sidebands cross $\pm 180°$ with loop gains over 1, indicating an unstable closed loop. The phase $\theta_k$ is used to tune the phase responses around the beam harmonic frequencies and keep the loop stable. The frequency response with $\theta_k = 50°$ in Fig. 17 shows that the loop phases at the beam harmonic frequencies are shifted towards $0°$, resulting in a stable closed loop. The $\theta_k$ of each harmonic controller can be optimized separately, e.g., proportional to the harmonic number $k$. With this method, the harmonic feedback controllers can suppress the beam loadings at frequencies far beyond the closed-loop bandwidth of the PI feedback controller. Figure 18 shows that up to 10 harmonic feedback controllers (still not at the limit) are successfully applied to the cavity simulator. The loop phases are set to be $\theta_k = k \times 20°$. The 10th beam harmonic frequency is 299.8 kHz, which is far beyond the PI control bandwidth around $K_p\omega_{1/2}$ (173 kHz).

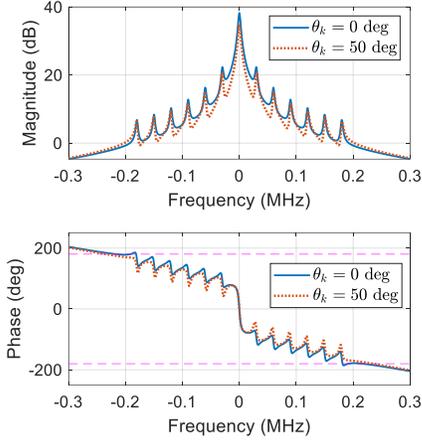

FIG 17. Frequency responses of **L** for different loop phases of harmonic feedback controllers.

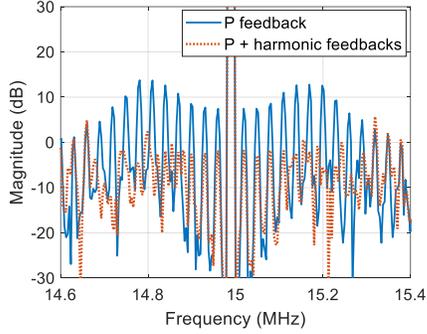

FIG 18. Cavity voltage spectrum with P feedback and 10 harmonic feedback controls.

In a practical RF system, the amplifiers' nonlinearity may significantly degrade the feedforward and feedback control performance. With nonlinearity, the feedforward calibration coefficients may become invalid after changing the RF operating point. This requires a re-calibration of the feedforward coefficients or adaptive algorithms. For feedback, the outputs of different harmonic controllers may crosstalk due to the intermodulation between different harmonic frequencies [5, 14], causing control instabilities. Therefore, RF amplifiers with high linearity should be chosen if the harmonic beam loadings or instabilities in circular accelerators should be critically controlled.

## V. BEAM LOADING IN CIRCULAR ACCELERATORS

The discussions beforehand are general for all cavities and beam loadings. In circular accelerators, particle bunches circulate for many turns, resulting in additional frequency components in the instant beam current, such as

- If the buckets of the circular accelerator are not evenly filled with bunches, or if the charge or phase of different bunches are not uniform, the instant beam current will contain harmonics of the bunch revolution frequency $\omega_{rev}$.
- Since the particles in a bunch are doing synchrotron oscillations, the harmonics of the synchrotron oscillation frequency $\omega_{syn}$ may also exit in the beam current spectrum if the phase-space distribution of the bunches have non-zero synchrotron oscillations.

Therefore, in a circular accelerator, the following frequency components may exist in the instant beam current experienced by the cavity:

$$\omega_{beam} = k\omega_b \pm n\omega_{rev} \pm m\omega_{syn} \qquad (24)$$

where $k$, $n$, $m$ are all non-negative integers satisfying $n < \omega_b/\omega_{rev}$ and $m < \omega_{rev}/\omega_{syn}$. The last two terms are typically related to beam instabilities such as the coupled-bunch instability [8-10]. When caused by the coupled-bunch instability, $n$ is the index of the coupled bunch mode and $m$ is the synchrotron oscillation mode (e.g., $m = 1$ corresponds to the dipole mode and $m = 2$ corresponds to the quadrupole mode, and so on).

The theory of bunch-train beam loading also applies to the beam in a circular accelerator possibly with beam instabilities. The frequency components of the instant beam current can be analyzed with the Fourier series, and then the induced cavity voltage can be calculated following the theory in Section III.B. The cavity response has contributions from both the nominal beam loading caused by a stable beam and the beam instability-caused frequency components. By counteracting the beam-induced cavity voltages, both the nominal beam loading and the beam instability can be suppressed, resulting in a stable cavity field.

The feedback controller (23) is fully applicable to stabilize the cavity field in a circular accelerator, with the harmonic feedback frequency $\omega_{n,k}$ chosen to be one of the frequencies in (24) that needs to be suppressed [15]. As mentioned before, if the synchrotron oscillation frequency changes over time, e.g., during the filling process of a storage ring equipped with harmonic cavities, the controller should be adapted to follow the changes of the disturbance frequency. The feedforward controller (22) can also be applied to suppress the beam instability in a circular accelerator. In this case, the frequency source of the feedforward controller should be taken from the beam, as Fig. 15. NCO is not suitable since the instability does not always exist or with time-varying frequencies, not like the nominal beam loading. The criteria to optimize the feedforward coefficients can either be the cavity signal spectrum at the corresponding frequencies or the beam instability directly measured by the beam monitor. When used for suppressing the beam instability, such beam-based feedforward algorithm is also known as the "mode-by-mode feedback" [16].

In the context of a global beam control system, the feedback and feedforward algorithms described in this article both use RF cavities as actuators. The control actions are applied to the RF system input, as shown in Fig. 10. The control bandwidth is strongly limited by the bandwidth of the main RF cavities used for beam acceleration. To control higher order instability modes, dedicated wideband cavities (also called "longitudinal kicker") may be adopted. The control algorithms are the same. With longitudinal kickers, the feedforward control scales and rotates the filtered beam signal to minimize the beam instability, and the feedback control measures the cavity field and keeps it zero by counteracting the beam induced fields.

The performance of the beam instability control can be evaluated by directly observing the spectrum of the beam monitor signal. However, since the beam loading effects exist for both stable and unstable beams, the beam loading control performance can only be evaluated by observing the spectrum of the cavity field signal. The same feedforward and feedback control algorithms given in this article are applicable to both control tasks.

## VI. CONCLUSIONS

In this work, the beam loading theory in standing-wave cavities is discussed systematically. Based on the PLT method, the cavity model and the responses to a single bunch and bunch train are derived. This method provides a general way to analyze the beam loading using the frequency components of the instant current beam. The typical feedforward and feedback algorithms are introduced for beam loading compensation. The algorithms are validated with a cavity simulator. Finally, the beam instability in circular accelerators is briefly discussed, indicating that the beam loading theory and control algorithms are also applicable to understand and control the beams with coupled-bunch instabilities.